# Superconductivity and Fast Proton Transport in Nanoconfined Water

K. H. Johnson*

Department of Materials Science and Engineering

Massachusetts Institute of Technology, Cambridge, MA 02139

A real-space molecular-orbital description of Cooper pairing in conjunction with the dynamic Jahn-Teller mechanism for high-$T_c$ superconductivity predicts that electron-doped water confined to the nanoscale environment of a carbon nanotube or biological macromolecule should superconduct below and exhibit fast proton transport above the transition temperature, $T_c \cong 230$ degK (-43 degC).

## 1. Introduction

A major goal of superconductor research and development is the discovery of useful substances that superconduct at the highest possible transition or critical temperatures, $T_c$. The Bardeen-Cooper-Schrieffer (BCS) theory of superconductivity,[1] which ascribes the onset of the superconducting state at the transition temperature $T_c$ to electrons attractively paired via virtual phonons, has been eminently successful in explaining conventional, relatively low-$T_c$ superconductors. However, BCS theory in its simplest form has failed to explain the origin of the high-$T_c$ superconductivity of doped cuprates.[2] Coupled with the dynamic Jahn-Teller (DJT) effect[3] and density-functional calculations for clusters representing the local molecular environments in superconducting materials,[4] a *real-space molecular-orbital* description of electronic wave functions which are precursors of the superconducting state in high- and low-dimensional metals was proposed and applied to a variety of superconductors.[5] According to this scenario, superconductivity is possible only if the normal chemical bonding system in the material or parts thereof permits the construction of vibronically-coupled degenerate or nearly-degenerate ("pseudo" Jahn-Teller) molecular-orbital wave functions at the Fermi energy ($E_F$) which, for at least one space direction, are not topologically intersected by plane or conical nodal surfaces.

---

*Professor Emeritus; kjohnson@mit.edu

This translates to the requirement of spatially delocalized molecular orbitals at $E_F$ that are bonding along "channels" of opposite phase, $\Psi_+$ and $\Psi_-$ which are a *coordinate-space* basis for the *Cooper-pair wavefunction*,

$$\Phi(\mathbf{r}) = <\Psi_+(\mathbf{r}+\mathbf{d}/2)\uparrow \Psi_-(\mathbf{r}-\mathbf{d}/2)\downarrow>, \qquad (1)$$

where $\mathbf{r}$ is the pair spatial vector and d is the distance between the $\Psi_+$ and $\Psi_-$ "channels," beyond which the electron-electron repulsion at $E_F$ is largely screened out by the intervening ion cores (see Figs. 2, 4, and 5).[5] The theory was applied to conventional metallic and organic superconductors,[5,6] high-$T_c$ cuprates,[6] and superconducting potassium-doped carbon fullerene,[7] yielding $T_c$-values, coherence lengths, isotope shifts, Debye frequencies, and thermodynamic critical magnetic fields in good agreement with experiments. Major findings include the association of cuprate high-$T_c$ superconductivity with DJT vibronically-coupled, mainly oxygen orbitals at $E_F$, the prediction of 230 degK (-43degC) as the upper limit of $T_c$, and reduction of the theory to conventional BCS theory in the limit of harmonic vibronic (electron-phonon) coupling.[5,6] In the present paper, we propose that *supercooled, nanoscopically confined, electron-doped* water should superconduct up to the highest predicted $T_c \cong 230$ degK due to DJT-induced terahertz (THz) vibrations of the component water nanoclusters coupled with their degenerate, mainly oxygen molecular orbitals at $E_F$. *Above* $T_c \cong 230$ degK, nanoconfined water should exhibit high proton conductivity.

## 2. Review of the Theory Applied to High-$T_c$ Cuprates and Fullerenes

Common to all the high-Tc cuprates is the approximately square-planar $CuO_4$ coordination complex forming the $CuO_2$ layers. Key to understanding both the parent insulating and superconducting phases of the cuprates is the strong covalency between the copper 3d atomic orbitals and the oxygen 2p valence orbitals, as compared with transition metals and oxygen. As a result, the relative ordering and characters of the Cu(d)-O(p) $\sigma$ and $\pi$ molecular-orbital levels in a $CuO_4$ coordination complex, determined from density-functional theory and shown in Fig. 1(b) are significantly different from those of the ligand-field levels for a typical ionic transition-metal complex $ML_4$ shown in Fig. 1(a). This is due to the strong $Cu(d_{xz,yz}\pi^*)$-$O(p_z\pi)$ covalent *antibonding* hybridization. It may be noted that some previous theories of high-$T_c$ superconductivity have assumed that the ordering of electron states in the doped $CuO_4$ complexes of the $CuO_2$ layers is the same as the $ML_4$ ligand-field levels of Fig. 1(a). That is simply incorrect. Occupancy of the strongly localized σ*-*antibonding* $b_{1g}$ $Cu(d_{x^2-y^2})$-$O(p_{x,y})$ orbital shown schematically in Fig. 1(b) is associated with the non-superconducting state of *undoped* cuprates. *Hole doping* leads to partial occupancy of the doubly degenerate $e_g$ $O(p_z)$-$Cu(d_{xz,yz})$ π*-*antibonding* molecula*r* orbital shown schematically in Fig. 1(b) and in the density-functional $e_g(xz)$ wavefunction contour map of Fig. 2. Note the predominant $O(p_z\pi)$ character. These coordinate-space $e_g$ orbitals correspond to the **k**-space *flat-band 4 and Van Hove singularity* at $E_F$ above the mainly oxygen-like valence band shown in the reproduced Fig 3(a).[8] The real-space $e_g(xz)$ molecular-orbital wavefunction contour map of Fig. 2 is very similar topologically to the charge-density map shown in Fig. 3(b) for this flat-band Van Hove singularity.[8] The key consequence of this strong $O(p_z)$-$Cu(d_{xz,yz})$ π*-antibonding hybridization at $E_F$ for optimally doped high-$T_c$ cuprates (not pointed out previously in publications by other authors) is the promotion of substantial $O(p_z\pi)$-$O(p_z\pi)$ bond overlaps of opposite phase, $\Psi_+$ and $\Psi_-$ *above and parallel to the CuO$_2$ layers*, as revealed in the contour map of the $e_g(xy)$ molecular-orbital wavefunction plotted 0.8 A above the $CuO_2$ layer in Fig. 2. The occupancy and $O(p_z\pi)$-$O(p_z\pi)$ bond overlaps within the $\Psi_+$ and $\Psi_-$ channels depend sensitively on the doping and further details of the electronic structure, such as the "puckering" of the $CuO_2$ layers and their $CuO_4$ coordination complexes, the presence of apical oxygens above and below the $CuO_2$ layer in $La_{2-x}Sr_xCuO_4$, and the presence of chains above and below the $CuO_2$ layers in $YBa_2Cu_3O_7$. The key point here is that *common to all* the optimally doped HTSC cuprates at $E_F$, there is significant $O(p_z\pi)$-$O(p_z\pi)$ bond overlap (equivalent to oxygen-oxygen hole "hopping") above and parallel to (but *not within*) the $CuO_2$ layers, forming the basis, $\Psi_+$ and $\Psi_-$ of a *Cooper-pair "density wave" above and parallel to each CuO$_2$ layer* in coordinate space as shown schematically in Fig. 4. A Cooper-pair density wave has indeed recently been detected in the cuprate, $Bi_2Sr_2CaCu_2O_{8+x}$.[9]

In the example of potassium-doped fullerene, icosahedral ($I_h$-symmetry) $C_{60}$ "buckyballs" are arranged in an fcc structure such that the electronic states at $E_F$ in real space are derived from degenerate $t_{1u}(p\pi)$ molecular orbitals that are bonding around the hemisphere of each buckyball and overlap, forming the *Cooper-pair basis wavefunctions,* $\Psi_+$ and $\Psi_-$ shown along one of three equivalent directions in Fig 5.[7] The interstitial $K^+$ ions that donate their valence electrons into the normally unoccupied buckyball $t_{1u}(p\pi)$ molecular orbitals are not shown. Because of the $C_{60}$ $t_{1u}(p\pi)$ molecular-orbital degeneracy and partial occupancy at modest potassium electron doping, each buckyball is subject to the DJT effect,[3] as is well known to occur in other pπ aromatic systems such as the benzene molecular ion.[10] Due to the modest overlaps of the $C_{60}$ pπ orbitals within the

$\Psi_+$ and $\Psi_-$ channels, the DJT $I_h$-symmetry-breaking vibrations ($H_g$ modes) of the buckyballs occur *coherently* and correspond to "soft-phonon" modes in the solid $K_xC_{60}$.

Based on the above molecular-orbital $\Psi_+$ and $\Psi_-$ topologies in both the doped cuprates and fullerenes, the Cooper pair can then be described as the *coordinate-space wavefuction*, Eq. 1. Including the symmetry-breaking effect of DJT vibronic (*local* electron-phonon) coupling of the degenerate $\Psi_+$ and $\Psi_-$ electronic states and solving the Schrödinger wave equation approximately for the Cooper-pair wavefunction, Eq.1 yields the following general practical formula for $T_c$,[5,6]

$$k_BT_c \cong h\nu_c\exp\{-h^2/2me^2d[1-(m/M)^\beta]\}, \qquad (2)$$

and DJT vibrational cut-off frequency,

$$\nu_c \cong h(m/M)^\beta/4\pi md^2, \qquad (3)$$

where $k_B$ is the Boltzmann constant, m is the electron mass, and M is the mass of the DJT-vibrating atom (oxygen mass number M = 16 in the cuprates and carbon mass number M = 12 in fullerene), The DJT vibronic coupling parameter, $\beta$ is determined from the respective computed O-O and $C_{60}$-$C_{60}$ $p\pi$-bond overlaps within the $\Psi_+$ and $\Psi_-$ channels (see Fig, 7a).[5-7] For the *"harmonic limit,"* $\beta = ½$, formula (2) is the real-space equivalent of BCS theory, and $\nu_c$ reduces to the Debye frequency.[5-7] For $¼ < \beta < ½$, the DJT effect corresponds to *anharmonic electron-phonon coupling*, moving $\nu_c$ into the terahertz (THz) range. For $\beta < ¼$, there is the possibility of reaching the highest $T_c$ of 230 degK (curve "e" of Fig. 6a), although the *static Jahn-Teller effect* can compete with the DJT effect and make the system structurally unstable, preventing Cooper pairing.[3,5] Indeed, the highest-$T_c$ materials have tended to lose their superconductivity because of structural changes induced by the static Jahn-Teller effect. Formulae (2) and (3) were applied to the high-$T_c$ cuprates[6] and potassium-doped fullerene,[7] yielding results in excellent agreement with experiments. Graphs of $T_c$ and isotope effect, $\propto = -\partial \ln T_c/\partial \ln M$ versus $\beta$ for oxygen mass number M = 16 and various values of the $\Psi_+$ - $\Psi_-$ interchannel distance, d displayed in Fig. 6 are representative of both the cuprates *and* of confined water (see Section 4), where oxygen is the DJT vibronically active element responsible for superconductivity. Note the vanishing of the isotope effect at the $T_c$ curve peaks in Fig. 6b. The relationship between anharmonicity, $\beta$ and $O(p\pi)$-$O(p\pi)$ bond overlap within the $\Psi_+$ and $\Psi_-$ channels (Figs. 2 and 4) is shown in Fig. 7a. All high-$T_c$ cuprates lie between curves b and c of Fig. 6, and the upper limit of $T_c \cong 230$ degK is predicted for oxides with $\Psi_+ - \Psi_-$ interchannel distance, $d \cong 8$ A. The present DJT mechanism sets an upper limit of $T_c \cong 230$ degK (-43 degC) for *any* potential superconductive material, as exemplified for hydrogen in Fig. 7b. $T_c$ graphs for $K_xC_{60}$ are similar. A report of sporadic superconductivity at $T_c = 230$ degK in one sample of the cuprate, $EuBa_2Cu_3O_{6+\delta}$ was published years ago.[11] Recently, hydrogen sulfide at high pressure has been observed to superconduct at $T_c = 203$ degK by an attributed anharmonic electron-phonon mechanism,[12] which could possibly be the DJT mechanism presented here.

## 3. Predicted Superconductivity of Nanoconfined Water Below $T_c \cong 230$ degK

In Fig.5, one can view fcc superconducting $K_xC_{60}$ (x = 3) as a system of "confined" $(C_{60})^{-3}$ buckyball clusters stabilized by the surrounding interstitial potassium ions $K^+$, which have donated one electron per $K^+$ ion into the otherwise lowest unoccupied, six-fold degenerate $t_{1u}(p\pi)$ molecular orbital (LUMO) of each $C_{60}$ cluster, subject to DJT vibronic (anharmonic electron-phonon) coupling. This observation and the recent experimental evidence for high-$T_c$ superconductivity in hydrogen sulfide ($H_2S$) at high pressure by an anharmonic electron-phonon mechanism[12] leads one to consider nanoconfined water ($H_2O$) as a possible superconductor. However, the limitation of $T_c \cong 230$ degK by the present theory would suggest only *supercooled* nanoconfined water as a possibility. Water can be rapidly supercooled to -43 degC (230 degK) – the upper limit to superconducting $T_c$ according to the present theory – or lower by confining water at nanoscale, *e.g.* in nanotubes.[13] Recent x-ray studies of water supercooled to -15 degC in a quartz capillary have revealed the presence of pentagonal dodecahedral and other pentagonal water clathrate nanoclusters similar to those shown in Figs. 8 and 10, which become dominant with decreasing temperature.[14] Therefore, it is expected that this trend would continue at even lower temperatures in a nanoscale environment. Pentagonal water clusters have been shown to be present in nanotubes[13] and proteins near hydrophobic amino acids,[15] where the hydrophobicity promotes the pentagonal clustering of water molecules. THz vibrations of water nanoclusters have been argued to be key to biomolecular function such as protein folding.[16]

Fig. 8 shows the computed density-functional ground-state molecular-orbital energies and vibrational modes of the *protonated* pentagonal dodecahedral water cluster, $(H_2O)_{21}H^+$ or $(H_2O)_{20}H_3O^+$. Like icosahedral $C_{60}$ buckyballs, pentagonal dodecahedral water clusters have their oxygens at the vertices of a dodecahedron having $I_h$ icosahedral symmetry. Of particular importance are the "squashing" and "twisting" vibrational modes of a pentagonal dodecahedral cluster shown in Fig. 9. Density-functional calculations for the pentagonal dodecahedral water cluster, $(H_2O)_{20}$, and larger water clusters yield the vibrational modes in Fig. 10. Common to all the water nanoclusters studied are: (1) lowest unoccupied (LUMO) energy levels like those in Fig. 8a, which correspond to the "S"-, "P"-, "D"- and "F"-like cluster "surface" orbital wavefunctions in Fig. 8b; and (2) bands of vibrational modes between 1 and 6 THz (Figs. 8 and 10) due to O-O-O "squashing" (or "bending") and "twisting" motions of the type in Fig. 9. The vectors in Figs. 8 and 10 represent the directions and relative amplitudes for the lowest THz modes corresponding to the O-O-O "bending" (or "squashing") motions of the water-cluster "surface" oxygen ions. Surface O-O-O bending vibrations of water clusters in this THz range have been observed experimentally.[17] When an extra electron is transferred to the LUMOs - the so-called *hydrated electron* – it is a bound state.[18] Thus, *electron doping* can put electrons into stable water-cluster molecular orbitals - *e.g.* the pπ orbitals of Figs. 8b and 11 – that are precursors to forming the $\Psi_+$ and $\Psi_-$ *coordinate-space Cooper-pair wavefunction*, Eq. 1, as exemplified in Figs. 4 and 5 respectively for a hole-doped cuprate and electron-doped fullerene. Moreover, like the cuprate high-$T_c$ scenario above (Figs. 2 and 4), it is the overlapping *oxygen* pπ orbitals surrounding the water nanocluster surfaces (Figs. 8b and 11) that should Cooper-pair (Eq. 1) via the DJT effect below $T_c \cong 230$ degK, provided the electron-doped, hydrated-electron water clusters are aggregated and supercooled to the solid state.

To accomplish this, one can confine water to *nanoscopic* environments such as carbon nanotubes,[13] silica capillaries,[14] or proteins,[15,16] where water can form nanoclusters responsible for the observed anomalously soft dynamics and anomalous quantum state of protons in nanoconfined water, including evidence for *anharmonic* intermolecular potentials and large-amplitude motions in nanotube water.[19,20] In the hydrophobic environment of a simple single-wall nanotube, water molecules can bond together to form clusters like the ones in Figs. 8-11, as shown in Fig. 12. Density-functional electronic-structure calculations for *electron-doped* water clusters of diameter 8 A (0.8 nm) confined to a 12 A (1.2 nm) single-wall carbon nanotube have been performed and yield overlapping hydrated-electron water-cluster "surface" pπ molecular orbitals for neighboring confined water clusters like those shown in Figs. 12 and 13. The latter are the basis, $\Psi_+$ and $\Psi_-$, for *coordinate-space Cooper pairing* in Eq. 1 via DJT vibronic coupling with the water-cluster oxygen surface THz-frequency vibrational modes like the ones shown by the vector amplitudes in Figs. 8 and 10, although there is also weak coupling of the water-cluster modes to THz vibrations of the nanotube walls.

While signs of superconductivity at only 0.55 degK in "ropes" of (presumably) "dry" carbon nanotubes have been observed,[21] there has been a recent claim of high-$T_c$ superconductivity in *water-treated graphite* powder.[22] The above formula (2) applied to nanotube-confined water clusters of diameter d = 8 A (0.8 nm) for a DJT coupling constant β = 0.11 – determined from the water-cluster O(pπ)-O(pπ) bond overlap (Figs. 7a and 13b)[5,6] - predicts a maximum $T_c \cong 230$ degK, like that for the cuprates (Fig. 6a, curve "e") – a very high $T_c$, but indicating *no possibility* of "room-temperature" superconductivity. Again, it must be emphasized that this prediction of superconductivity in nanotube-confined water depends on the effective *electron doping* of the water, which can be done in the laboratory is several ways.

## 4. Fast Proton Transport in Carbon-Nanotube-Confined Water Above $T_c \cong 230$ degK

If nanotube-confined water clusters are *protonated*, like the ubiquitous $(H2O)_{21}H^+$ cluster shown in Fig. 8, and then electron-doped, *i.e.* electrons are added to the LUMOs, occupation of the LUMO cluster molecular orbitals shown in Figs. 8a and 8b will induce the symmetry-breaking dynamical Jahn-Teller effect (Fig. 9), which causes each confined water nanocluster (Fig. 13a,b) to oscillate *anharmonically* between "double-well potentials" (Fig. 13c) along the nanotube, lowering the energy barrier, $E_{barrier}$ for proton tunneling between neighboring water clusters. This is consistent with the observed anomalously soft dynamics and anomalous quantum state of protons in nanoconfined water, including evidence for *anharmonic* intermolecular potentials and large-amplitude motions in nanotube-confined water.[19,20] This promotes fast transport of protons through nanotube-confined water, as recently observed experimentally in sub-1-nm diameter carbon nanotube porins.[23] The phase-change transition temperature from superconducting Cooper pairs, Eq. 1, to highly conducting protons is given by Eq. 2 (the same $T_c$ formula for the upper limit of electronic superconductivity) because that phase change can be viewed as a transition from the *dynamic* to *cooperative* Jahn-Teller effect.[3] For confined 0.8-nm-diameter water nanoclusters like those shown in Fig. 12, $T_c \cong 230$ degK. Using Eq. 2, $\nu_c \cong 29$ THz, which is close to the upper-frequency limits shown in Figs. 8c and 10 for isolated water-cluster O-H "librational modes" which should promote proton transport *above* $T_c \cong 230$ degK, whereas *below* $T_c$, the watercluster mainly oxygen "surface vibrations" should promote superconductivity.

## 5. Fast Proton Transport in Biologically-Confined Water Above $T_c \cong 230$ degK

Another example of nanoconfined water is hydrated *phycocyanin*, the active center of an important light-harvesting protein,[24] where high electrical conduction possibly due to fast proton hopping has been experimentally observed to occur from -40 degC to room temperature, in good agreement with the above predicted $T_c$ value, -43 degC (230 degK) for fast proton transport.[25] First-principles molecular-dynamics calculations and experiment[25] reveal the formation of water nanoclusters within the phycocyanin cavities as shown in Fig. 14a. Density-functional calculations indicate the presence of degenerate water-cluster "pπ" molecular orbitals like the ones shown in Fig. 14b and Fig. 11 that are subject to the cooperative JT effect and its promotion of proton transport as described above. It is also possible that *below* $T_c \cong 230$ degK, these orbitals form the basis, $\Psi_+$ and $\Psi_-$ for coordinate-space Cooper pairing and therefore possible superconductivity of biologically-confined water as predicted above for nanotube-confined water.

## 6. Summary and Conclusions

Almost thirty years since the discovery of the high-$T_c$ cuprates there is still no consensus on the origin of its unusual superconductivity beyond associating it principally with the oxygen ions. Although "non-phonon" mechanisms for the cuprates have been popular over the years, the present, universal *"real-space"* theoretical model for superconductors, first proposed over thirty years ago[5] and applied with some success to the cuprate and fullerene superconductors,[6,7] is most consistent with "**k**-space" electron-phonon mechanisms for the cuprates based on van Hove singularities in the flat, mainly oxygen band at $E_F$.[8,26,27] (see Fig. 3a). The recent observation of a "real-space" Cooper-pair density wave in $Bi_2Sr_2CaCu_2O_{8+x}$ may add some credibility to the present DJT real-space molecular-orbital scenario.[9] Adapting this scenario to nanoconfined water, it is predicted that DJT-induced "water-nanocluster surface" (mainly oxygen) vibrations (anharmonic local phonons) in the 1-6 THz range (Figs. 8 and 10) should promote *hydrated-electron* Cooper pairing and superconductivity of the nanoconfined water below $T_c \cong 230$ degK, whereas above this $T_c$, the confined water-nanocluster "librational" modes (mainly hydrogen librations) in the 10-32 THz range (Figs. 8 and 10) should dominate and promote fast proton conduction. While there is indeed recent experimental evidence for fast proton transport in nanotubes[23] and the protein, phycocyanin,[25] it remains to be proven that such nanoconfined water will also electronically superconduct below $T_c \cong 230$ degK, as predicted in this paper.

**7. Acknowledgements**

The author thanks M. Price-Gallagher, Y. Feldman, and Y. Paltiel for helpful discussions.

**Figure Captions**

**Fig. 1.** (a) Ligand-field electron states of a typical square-planar transition-metal complex, $ML_4$. (b) Molecular-orbitals of a square-planar $CuO_4$ cluster.

**Fig. 2.** Wavefunction contour maps of the $CuO_4$ cluster $e_g(xy)$ and $e_g(xz)$ molecular-orbitals.

**Fig. 3.** (**a**) Fig. 1 of Ref. 8. $YBa_2Ba_2Cu_3O_7$ energy bands near $E_F$. (**b**) Fig. 5 of Ref 8. Charge-density counter map for band 4 of (**a**). Compare with $e_g(xz)$ molecular orbital in Fig. 2.

**Fig. 4.** The degenerate, coordinate-space, molecular-orbital basis, $\Psi_+$ and $\Psi_-$ of the Cooper-pair wavefunction, Eq. 1 above and parallel to each $CuO_2$ layer of a typical high-$T_c$ cuprate superconductor.

**Fig. 5.** The coordinate-space basis, $\Psi_+$ and $\Psi_-$ of the Cooper-pair density wave for a potassium-doped fullerene superconductor.

**Fig. 6.** Graphs of superconducting transition temperature, $T_c$ and isotope effect, $\propto$ versus dynamic Jahn-Teller coupling parameter, $\beta$ for oxygen mass number $M = 16$ and various values of Cooper-pair $\Psi_+$ - $\Psi_-$ interchannel distance d according to Eq. (2).

**Fig. 7. a.** Graph showing the relationship of $O(p\pi)$-$O(p\pi)$ bond overlap within the Cooper-pair basis $\Psi_+$ and $\Psi_-$ channels and the DJT coupling parameter, $\beta$. **b.** Graph of superconducting transition temperature, $T_c$ versus dynamic Jahn-Teller coupling parameter, $\beta$ for hydrogen and various values of the Cooper-pair $\Psi_+$ - $\Psi_-$ interchannel distance, d according to Eq. (2).

**Fig. 8.** Ground-state density-functional molecular-orbital states and vibrational modes of an $(H_2O)_{21}H^+$ protonated pentagonal dodecahedral water nanocluster. **a.** Cluster molecular-orbital energy levels. **b.** Wavefunctions of the lowest unoccupied cluster molecular orbitals. **c.** Vibrational spectrum. **d.** The 1.5 THz vibrational mode. The vectors show the directions and relative amplitudes of the O-O-O "bending" oscillations of the cluster "surface" oxygen atoms.

**Fig. 9.** "Squashing" and "twisting" vibrational modes of a pentagonal dodecahedron. Hg and Hu designate the key irreducible representations of the icosahedral point group corresponding to these modes.

**Fig. 10**. **a.** Vibrational spectrum of a pentagonal dodecahedral $(H_2O)_{20}$ water cluster. **b.** Lowest-frequency THz vibrational mode of the cluster. **c.** Vibrational spectrum of an array of three dodecahedral water clusters. **d.** Lowest-frequency THz vibrational mode. **e.** Vibrational spectrum of an array of five dodecahedral water clusters. **f.** Lowest-frequency THz vibrational mode. The vectors show the directions and relative amplitudes for the O-O-O "bending" motions responsible for the "squashing" mode of the cluster "surface" oxygen atoms.

**Fig. 11.** Hydrated-electron "pπ" molecular-orbital wavefunction of a water nanocluster consisting of three connected pentagonal dodecahedra. See Fig. 3c,d for THz vibrations thereof.

**Fig. 12. (a)** Hydrated-electron "pπ" molecular-orbital wavefunction of a water nanocluster of diameter 0.8 nm (8 A) confined to a single-wall 1.2-nm-diameter carbon nanotube saturated at its ends with hydrogen. **(b)** Hydrated-electron "pπ" molecular-orbital wavefunction of a water nanocluster of diameter 0.8 nm confined to a single-wall 1.2-nm-diameter carbon nanotube.

**Fig. 13. (a)** Hydrated-electron "pπ" molecular-orbital wavefunctions of pentagonal dodecahedral water clusters of diameter 0.8 nm confined to an array of 1.2-nm-diameter single-wall carbon nanotubes (displayed end-on) labeled with the $\Psi_+$ and $\Psi_-$ components of the Cooper-pair wavefunction. **(b)** Overlapping hydrated-electron "pπ" molecular-orbital wavefunctions of nanoconfined neighboring pentagonal dodecahedral water clusters labeled with the $\Psi_+$ and $\Psi_-$ components of the Cooper-pair wavefunction. **(c)** Double potential energy wells for Jahn-Teller distorted water pentagonal dodecahedral water clusters (Fig. 9) and the resulting reduction of the energy barrier to proton transport between nanotube-confined neighboring water nanoclusters.

**Fig 14. (a)** Structure of the light-harvesting protein, phycocyanin confining a water nanocluster. **(b)** Hydrated-electron "pπ" molecular-orbital wavefunction of the phycocyanin-confined water nanocluster.

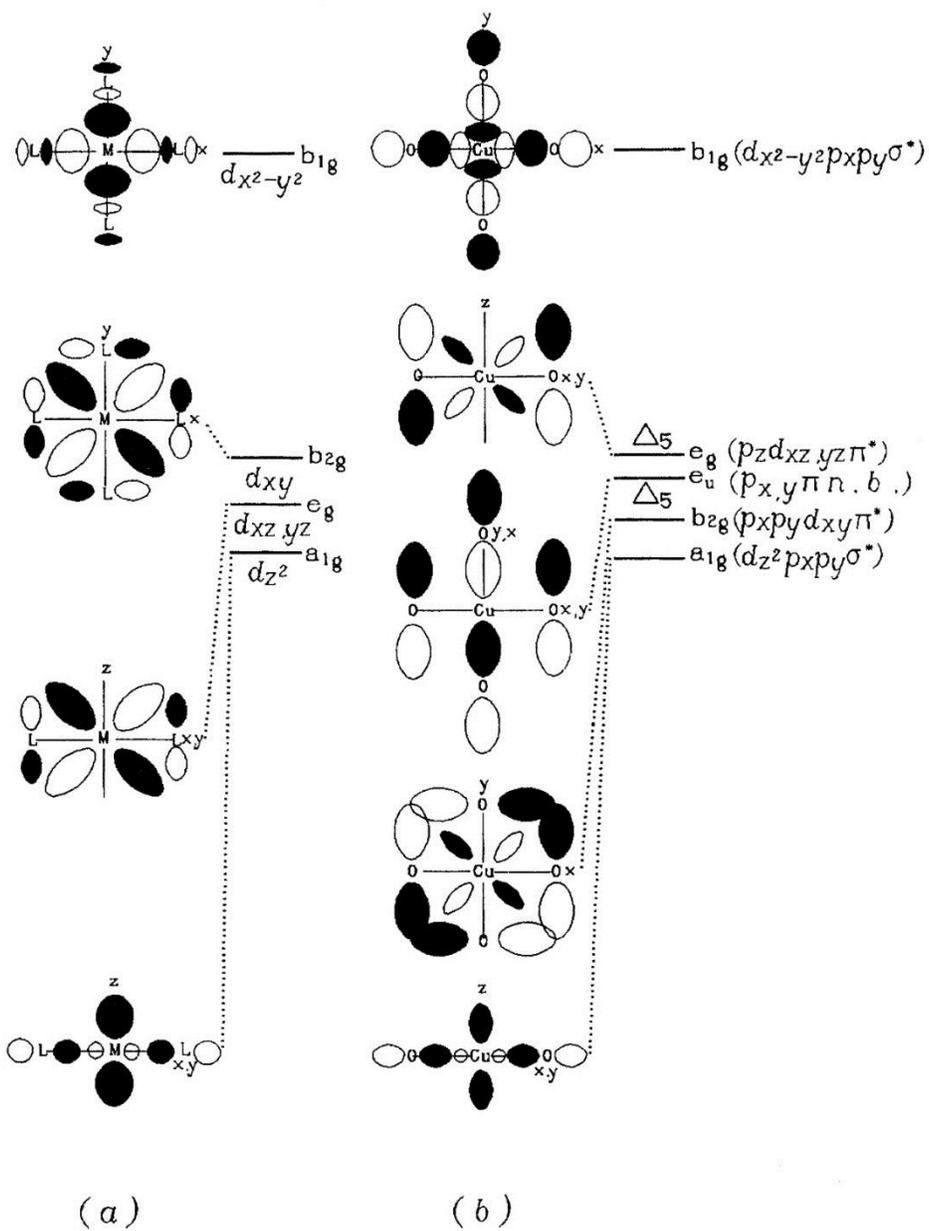

(a)     (b)

**Fig. 1**

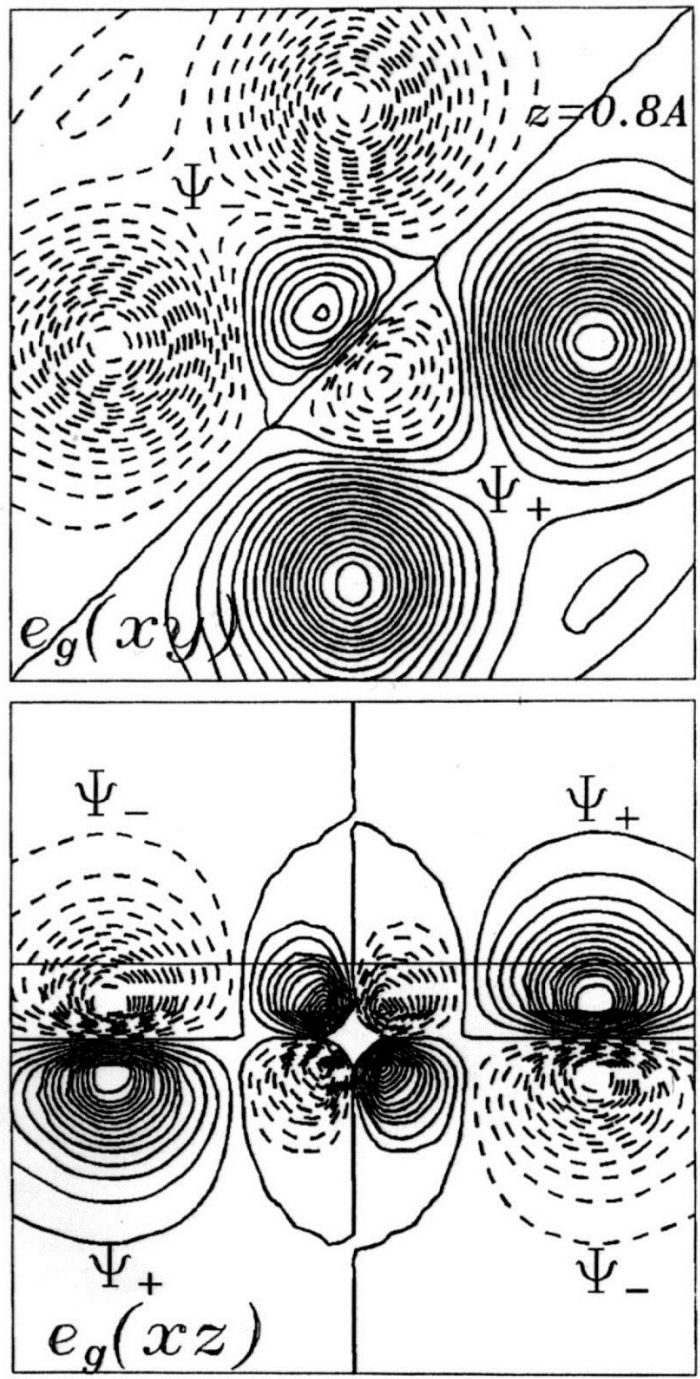

**Fig. 2**

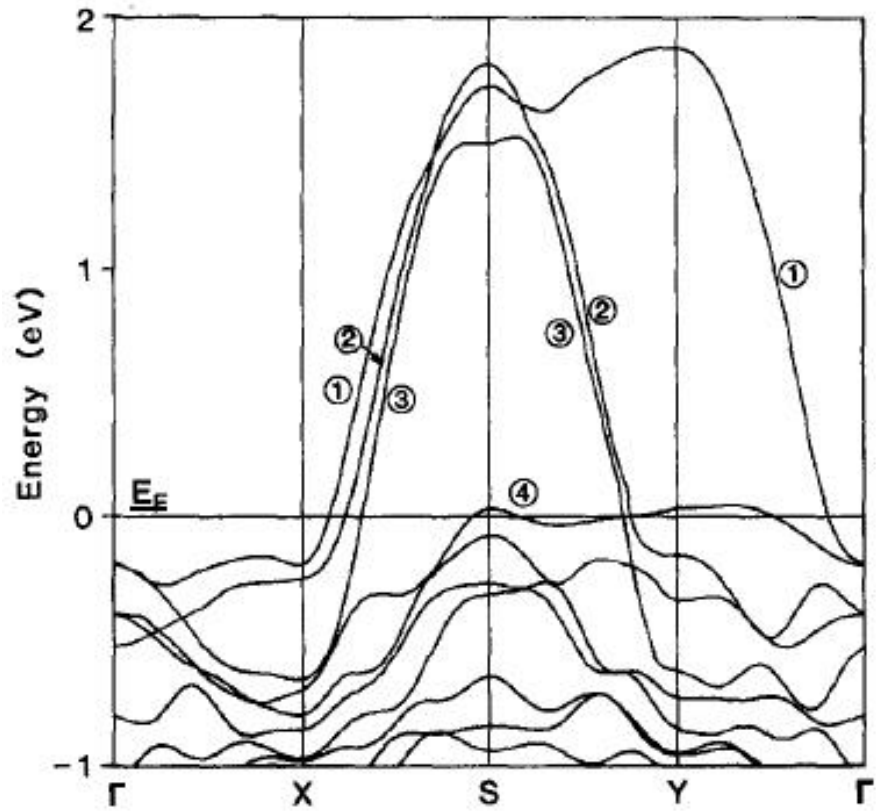

(a)

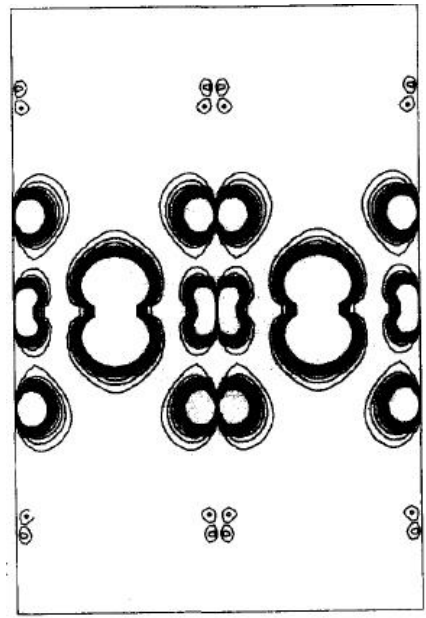

(b)

Fig. 3

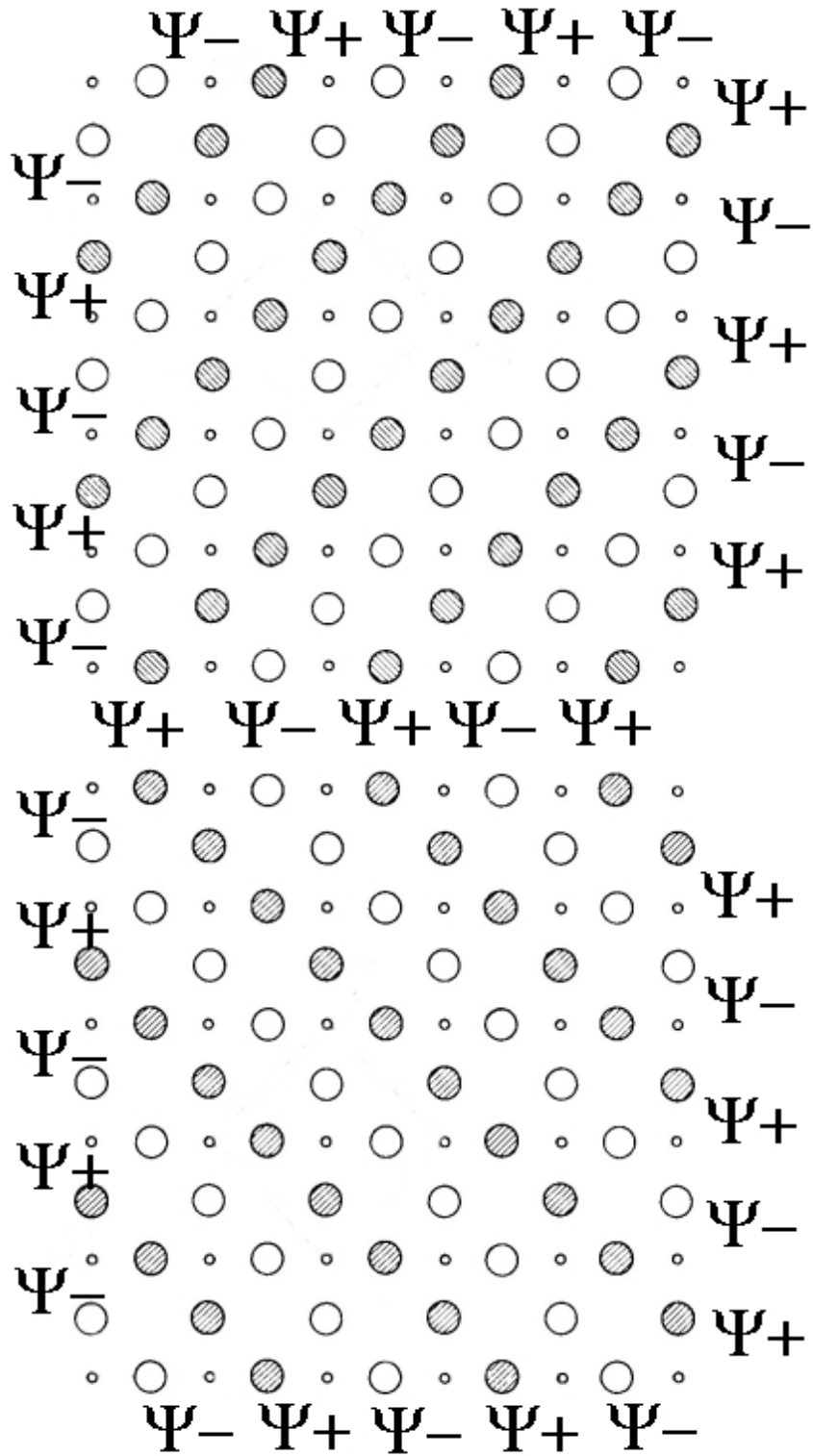

Fig. 4

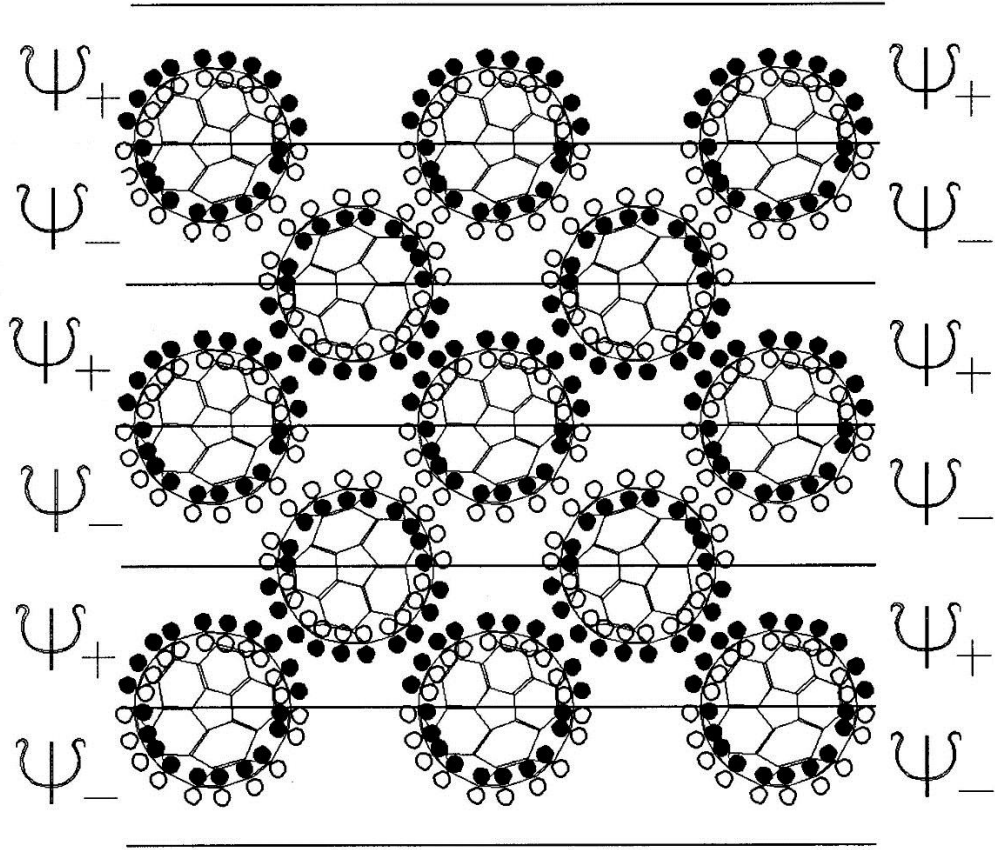

**Fig. 5**

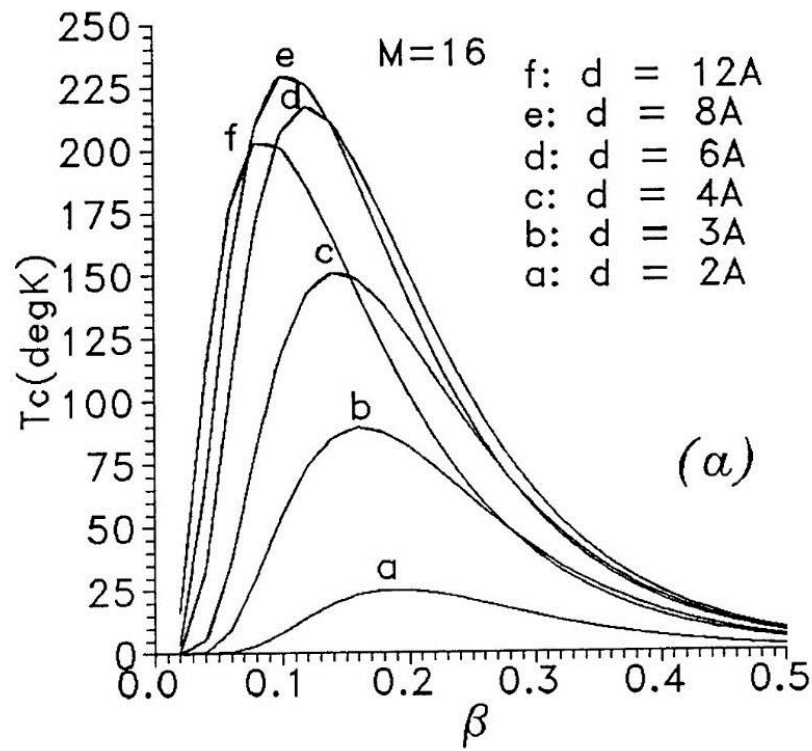
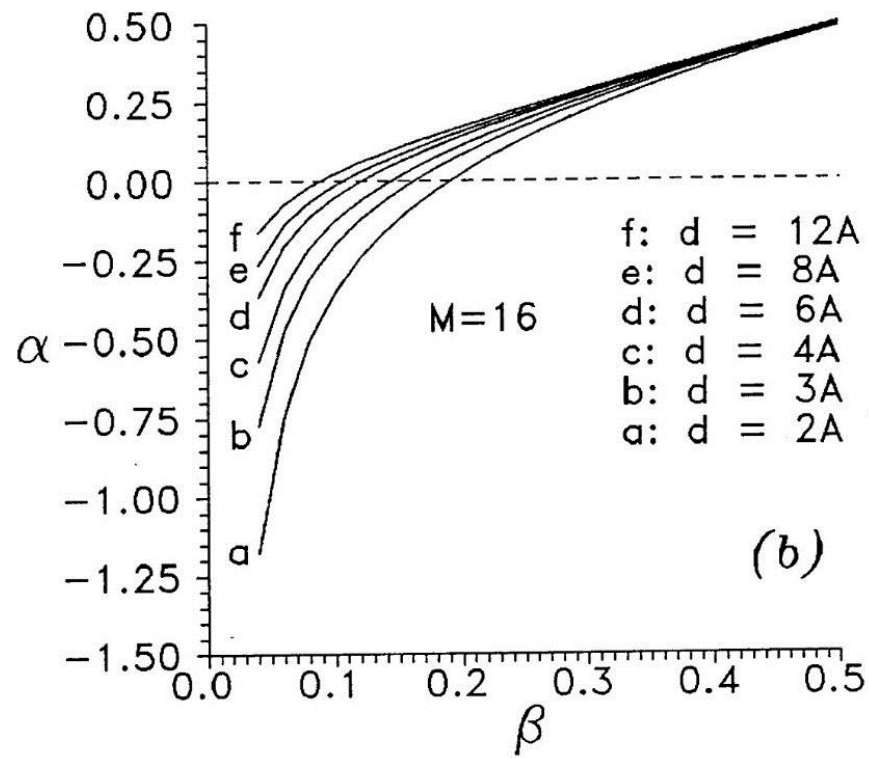

**Fig. 6**

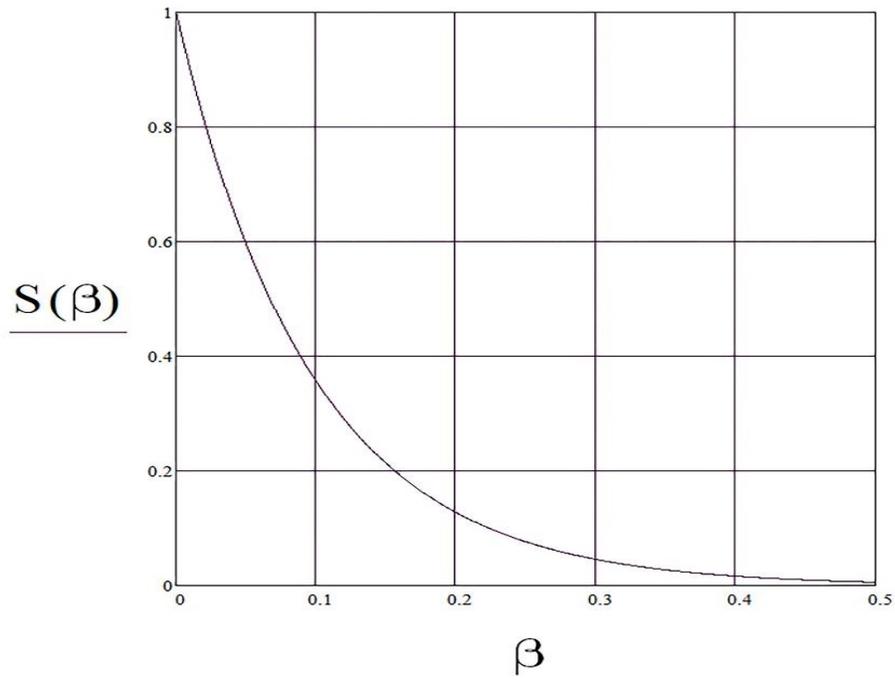

**(a)**

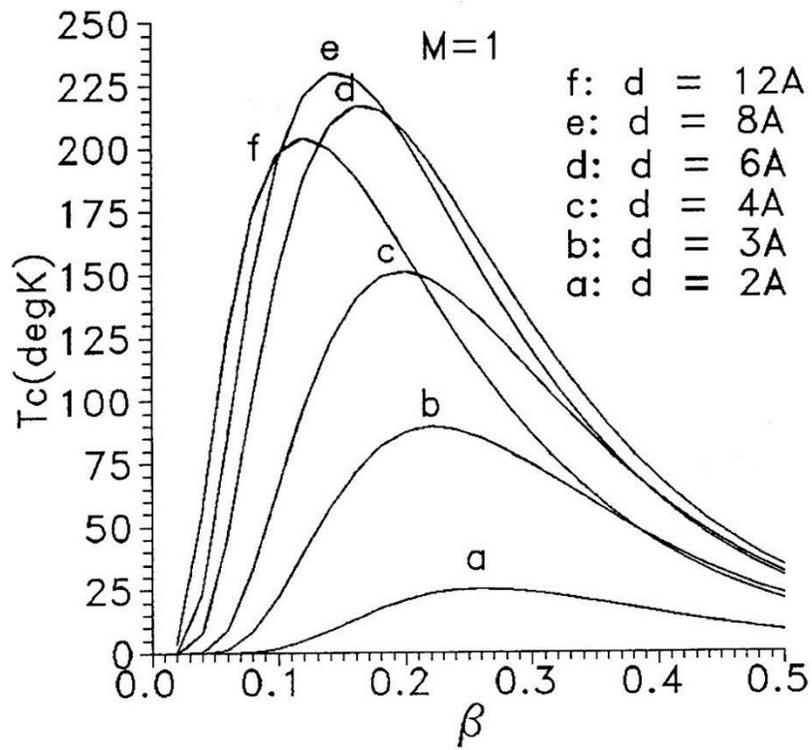

**(b)**

**Fig. 7**

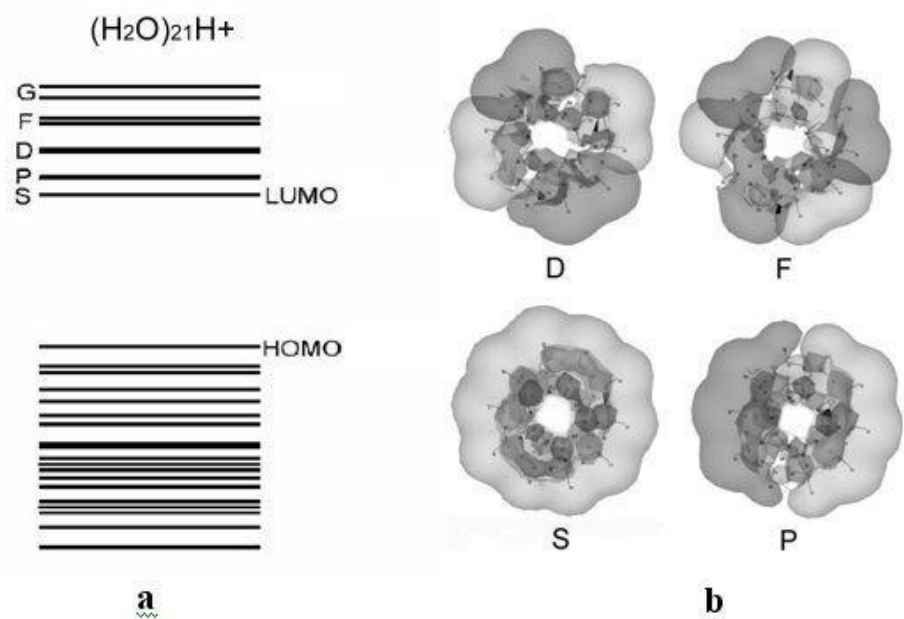
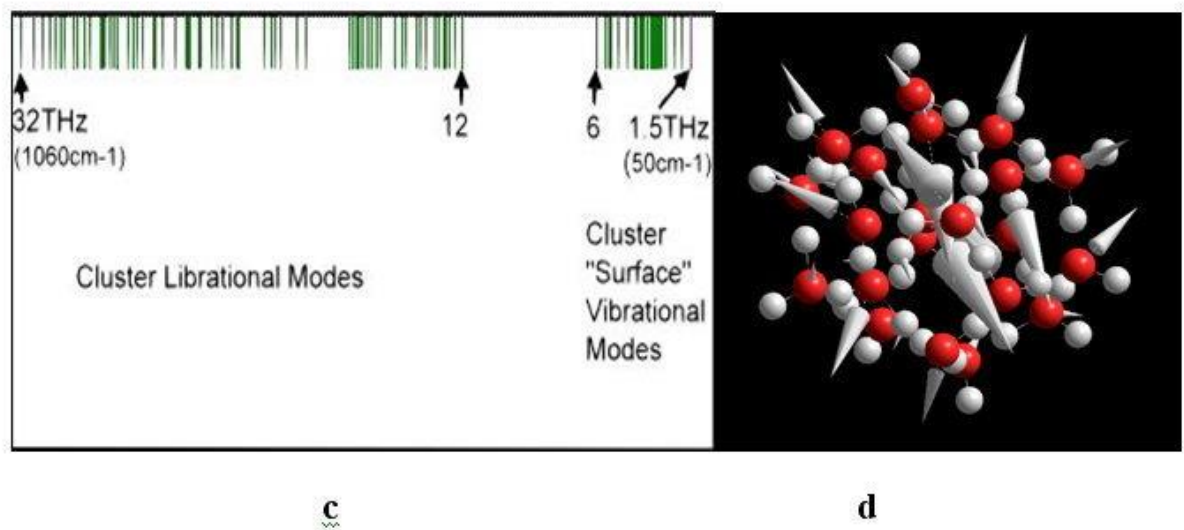

**Fig. 8**

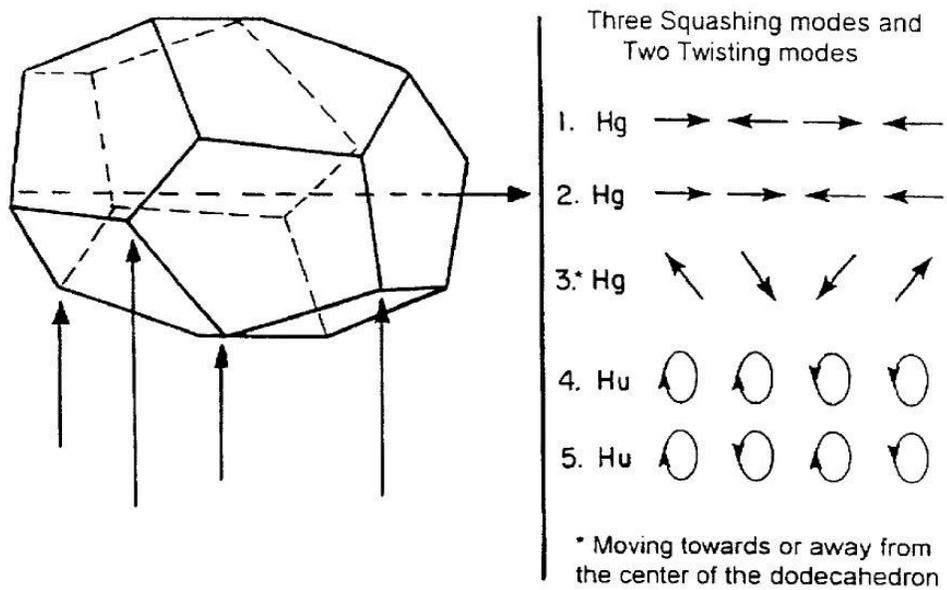

**Fig. 9**

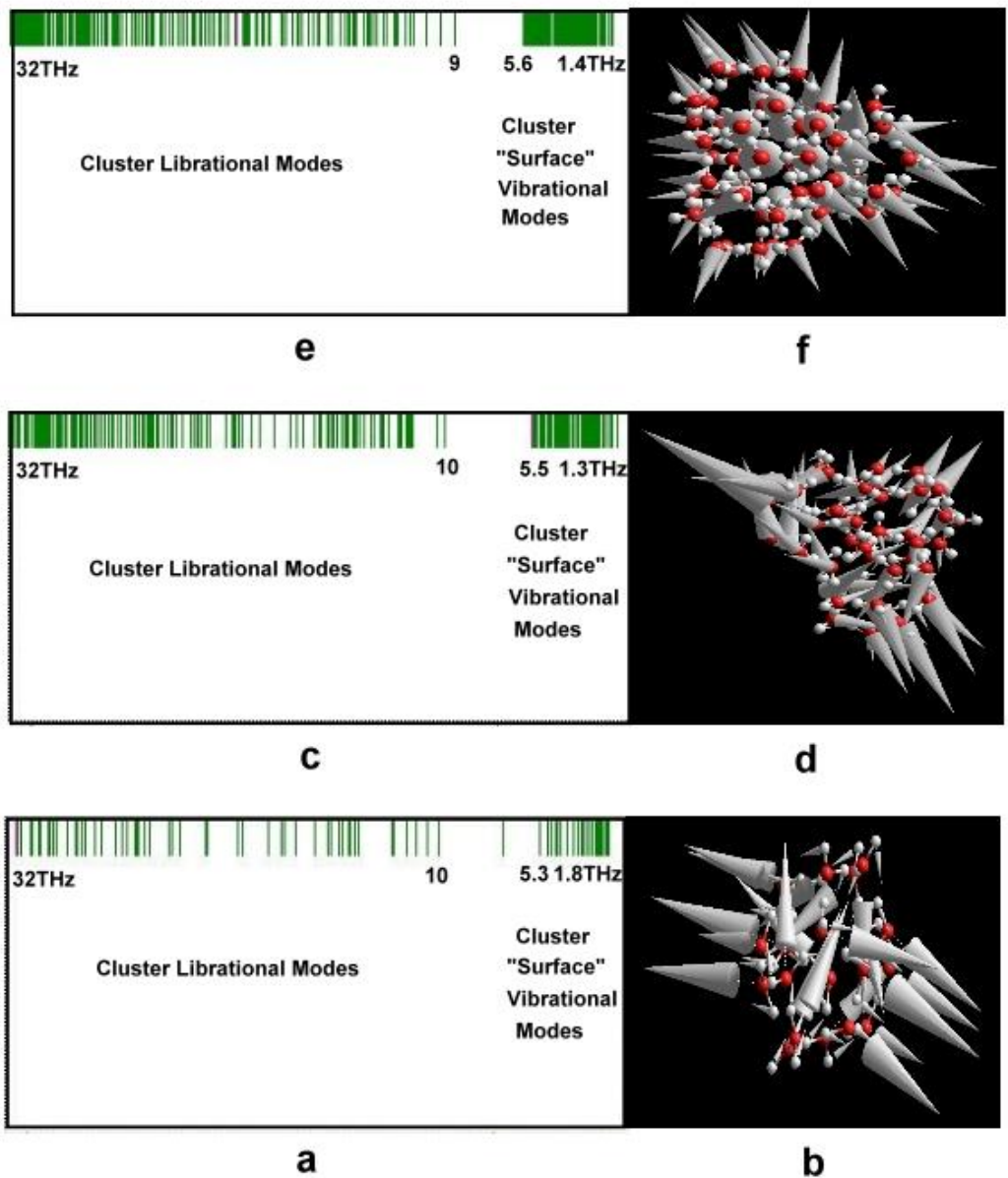

Fig. 10

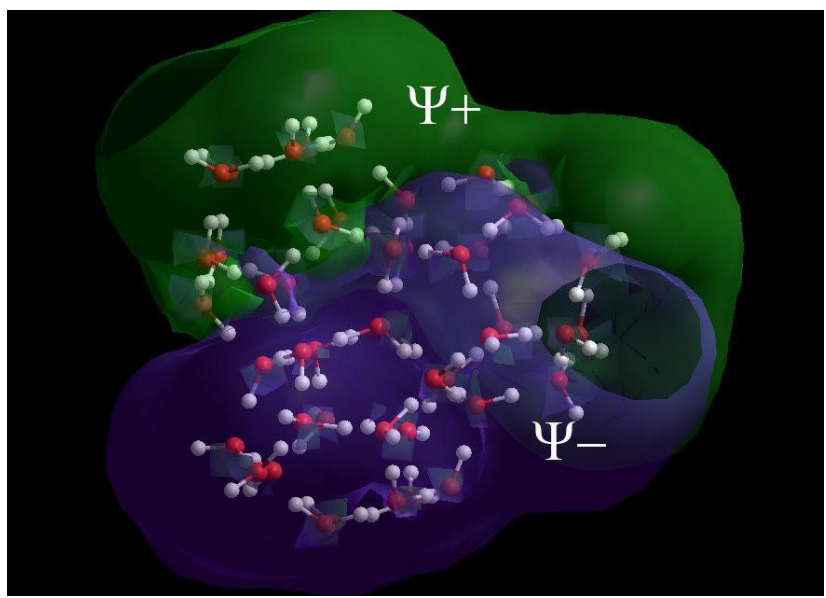

**(a)**

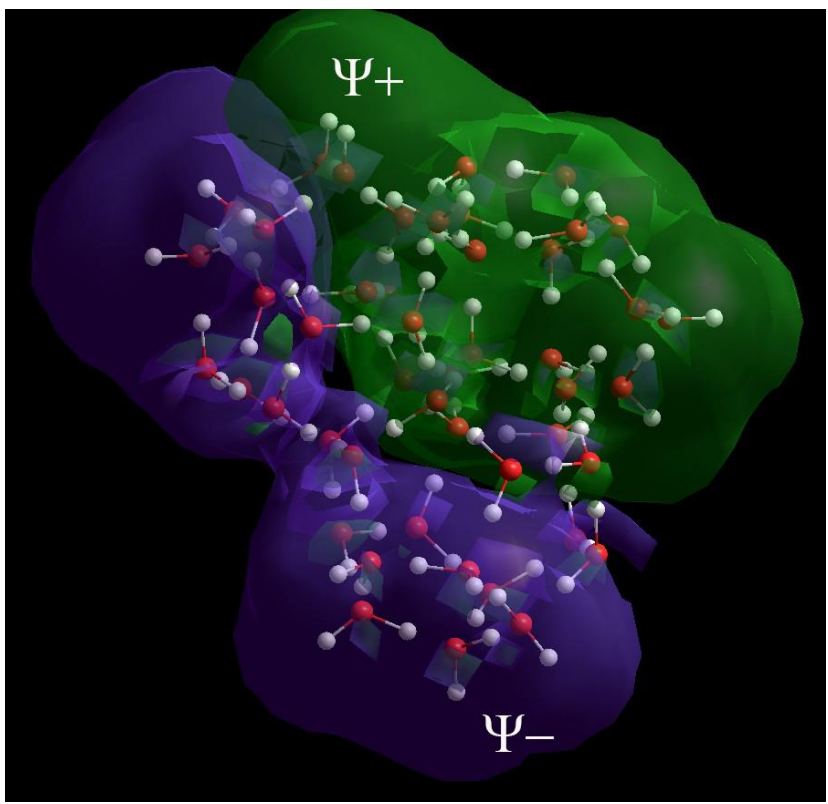

**(b)**

**Fig. 11**

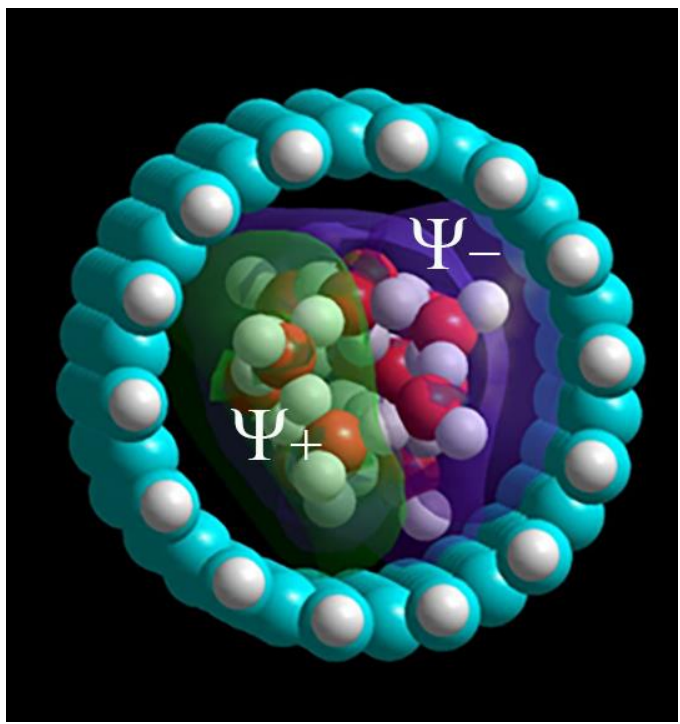

**(a)**

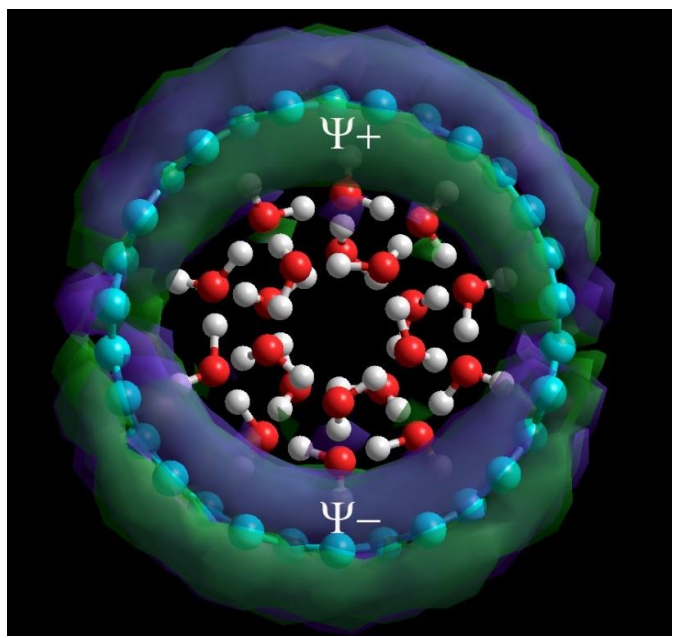

**(b)**

**Fig. 12**

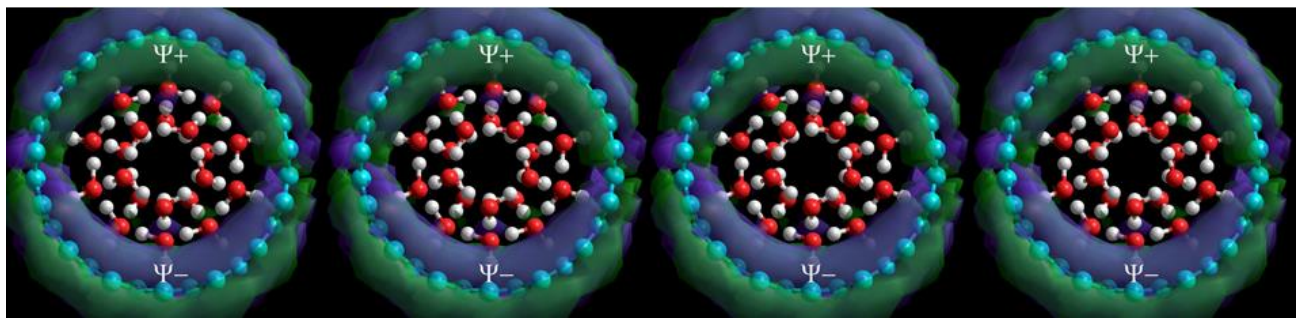

(a)

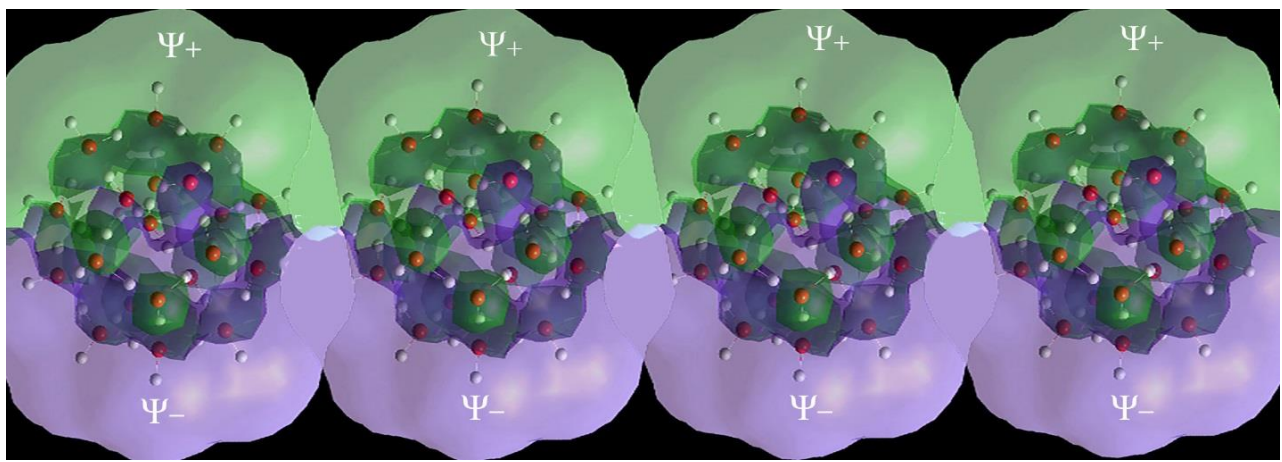

(b)

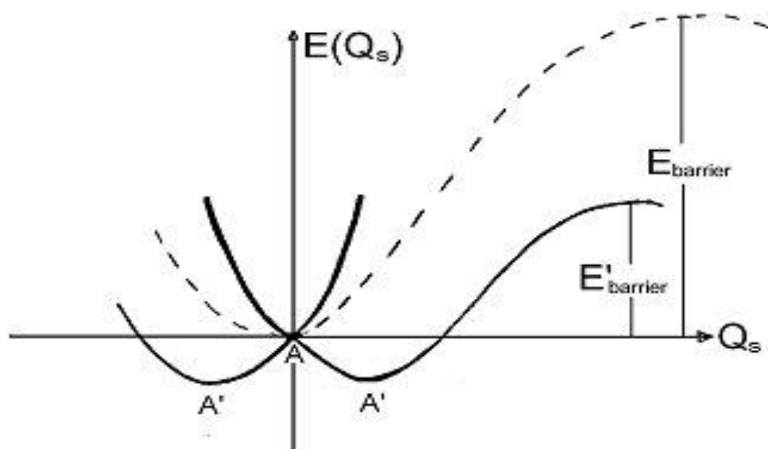

(c)

**Fig. 13**

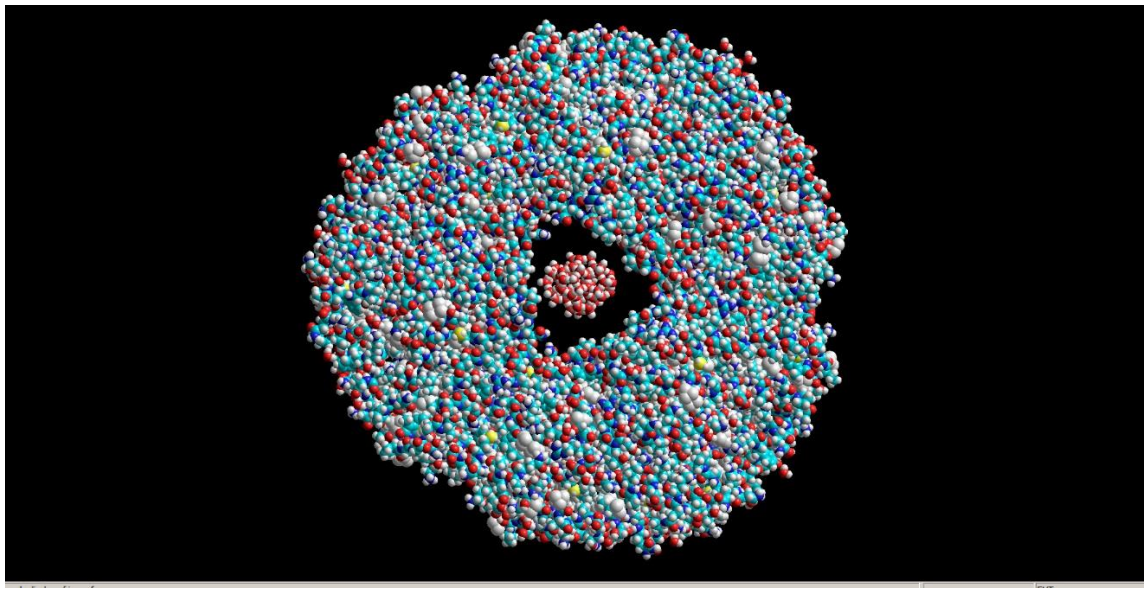

**(a)**

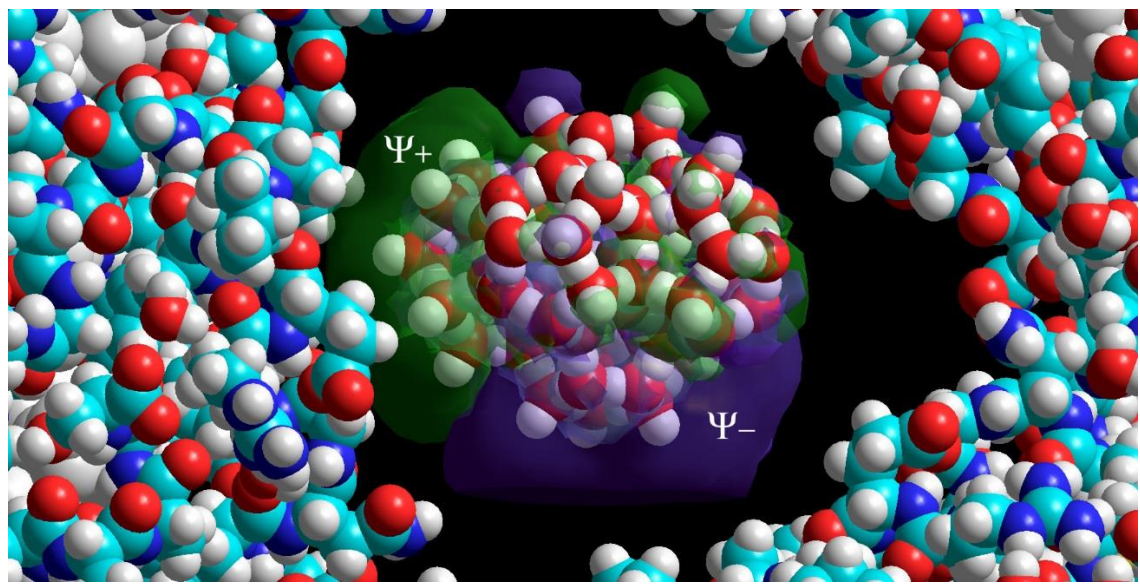

**(b)**

**Fig. 14**